\documentstyle[11pt,amssymb]{article}
\newcommand{\be}{\begin{equation}}
\newcommand{\ee}{\end{equation}}
\newcommand{\ba}{\begin{array}}
\newcommand{\ea}{\end{array}}
\newcommand{\bea}{\begin{eqnarray}}
\newcommand{\eea}{\end{eqnarray}}

\renewcommand{\l}{\newline\null}

\newcommand{\p}{\partial}
\newcommand{\ol}{\overline}

\newcommand{\la}{\langle}
\newcommand{\ra}{\rangle}

\def\hbar{h\!\!\!/}

\begin{document}
\begin{titlepage}
June 1996 \hfill PAR-LPTHE 96/20
\vskip 4cm
{\centerline{\bf CHIRAL SCALAR FIELDS,}}
{\centerline{\bf CUSTODIAL SYMMETRY IN ELECTROWEAK 
                                     {\boldmath $SU(2)_L\times U(1)$}}}
{\centerline{\bf AND THE QUANTIZATION OF THE ELECTRIC CHARGE.}}
\vskip .5cm
\centerline{B.~Machet
     \footnote[1]{Member of `Centre National de la Recherche Scientifique'.}
     \footnote[2]{E-mail: machet@lpthe.jussieu.fr.}
           }
\vskip 5mm
\centerline{{\em Laboratoire de Physique Th\'eorique et Hautes Energies,}
     \footnote[3]{LPTHE tour 16\,/\,1$^{er}\!$ \'etage,
          Universit\'e P. et M. Curie, BP 126, 4 place Jussieu,
          F 75252 PARIS CEDEX 05 (France).}
}
\centerline{\em Universit\'es Pierre et Marie Curie (Paris 6) et Denis
Diderot (Paris 7);} \centerline{\em Unit\'e associ\'ee au CNRS D0 280.}
\vskip 1.5cm
{\bf Abstract:}  I study the scalar representations of the electroweak 
group of the Standard Model, which is a subgroup of the chiral group 
$U(N)_L \times U(N)_R$ with $N$ flavours, for $N$ even, 
with a special emphasis on their chiral
properties and on their behaviour by the discrete symmetries 
$P$ and $CP$. They exhaust the $2N^2$ scalar and pseudoscalar degrees of 
freedom of the chiral group, for which a $SU(2)_L \times U(1)$  renormalizable,
anomaly-free gauge theory  naturally springs out.
It is shown to have a global diagonal $SU(2)_V$ symmetry independently of
the value of the hypercharge coupling, which becomes
local at the limit when the latter vanishes.
When acting in the 4-dimensional space of states spanned by the special
representations under scrutiny in this paper, 
the electric charge is one of the three generators of this 
``custodial''  symmetry; that the latter stays an unbroken symmetry is thus 
correlated with the quantization of the electric charge.
\smallskip

{\bf PACS:} 02.20.-a, 11.15.-q, 11.30.Er, 11.30.Rd, 12.15.-y 
\vfill
\end{titlepage}
\section{Introduction.}\label{sec:INTRO}

While scalar fields play a crucial and ambiguous role in the spontaneously
broken gauge theory of electroweak processes \cite{GlashowSalamWeinberg},
pseudoscalar and scalar
mesons are traditionally attached to the chiral group of strong
interactions \cite{CurrentAlgebra}. 
However, the dominance of their electroweak interactions 
makes mandatory their description within the framework of a gauge theory too.
This is achieved in this paper, which also re-unites within the same
framework the fields at the origin of the breaking of the symmetry and the
observed particles (mesons).

I reduce the most general $J=0$ representations of the electroweak
group to  $N^2/2$ quadruplet representations. They can be classified
according to their transformations by $CP$.

The existence of a quadratic invariant for those representations 
enables to write of a $SU(2)_L \times U(1)$ gauge invariant Lagrangian 
for $J=0$ mesons,  which is renormalizable by power counting.
It has furthermore a global diagonal $SU(2)_V$ symmetry, which
becomes  local at the limit $g' \rightarrow 0$, 
where $g'$ is the hypercharge $U(1)$ coupling constant.

In the space of states spanned by these representations, the electric charge 
is one of the three generators of the latter ``custodial'' $SU(2)_V$.
Its quantization, as that of the third component of an angular momentum,
is thus correlated with this $SU(2)_V$ being an exact symmetry of
the model.
It is to be related, from the point of view of electric-magnetic duality
\cite{Olive}, to a recent paper by Cho and Maison \cite{ChoMaison} uncovering
dyon-type classical solutions in this model.

Only briefs remarks concerning the phenomenology of observed electroweak
eigenstates  are made here.

\section{The chiral group  \boldmath{$U(N)_L \times U(N)_R$}.}
\label{sec:CHIRAL}

A generator $\Bbb A$ of $U(N)_L \times U(N)_R$ is a set of two $N\times N$
matrices $({\Bbb A}_L, {\Bbb A}_R)$. A generator of a diagonal subgroup
satisfies ${\Bbb A}_L =  {\Bbb A}_R$.

Both left and right parts of the chiral group violate parity; hence it is 
natural to classify the $J=0$ fields according to their behaviour by the 
parity changing operator $\cal P$, which transforms a scalar into a
pseudoscalar and vice-versa; we shall accordingly consider the action of 
the chiral group on $\cal P$-even or $\cal P$-odd states.

We define it  by the actions of its left and right commuting  subgroups. 
At the level of the algebra:
\bea
{\Bbb A}^i_L\,.\, {\Bbb M}_{{\cal P}even} &\stackrel{def}{=}&
            - \,{\Bbb A}^i_L \,{\Bbb M}_{{\cal P}even}
                                           ={1\over 2} \left(
            [{\Bbb M}_{{\cal P}even},{\Bbb A}^i_L]
                            - \{{\Bbb M}_{{\cal P}even},{\Bbb A}^i_L\}
                                    \right),\cr
{\Bbb A}^i_L\,.\, {\Bbb M}_{{\cal P}odd} &\stackrel{def}{=}& 
            +\,{\Bbb M}_{{\cal P}odd}\,{\Bbb A}^i_L
                                           =\hskip 5mm {1\over 2} \left(
            [{\Bbb M}_{{\cal P}odd},{\Bbb A}^i_L]
                            + \{{\Bbb M}_{{\cal P}odd},{\Bbb A}^i_L\}
                                    \right),\cr
{\Bbb A}^i_R\,.\, {\Bbb M}_{{\cal P}even} &\stackrel{def}{=}& 
            +\,{\Bbb M}_{{\cal P}even} \,{\Bbb A}^i_R
                                           ={1\over 2} \left(
            [{\Bbb M}_{{\cal P}even},{\Bbb A}^i_R]
                            + \{{\Bbb M}_{{\cal P}even},{\Bbb A}^i_R\}
                                    \right),\cr
{\Bbb A}^i_R\,.\, {\Bbb M}_{{\cal P}odd} &\stackrel{def}{=}& 
            - \,{\Bbb A}^i_R \,{\Bbb M}_{{\cal P}odd}
                                          =\hskip 5mm{1\over 2} \left(
            [{\Bbb M}_{{\cal P}odd},{\Bbb A}^i_R]
                            - \{{\Bbb M}_{{\cal P}odd},{\Bbb A}^i_R\}
                                    \right),
\label{eq:actionUNa}
\eea
which is akin to left- and right- multiplying $N\times N$ matrices.

{From} eqs.~(\ref{eq:actionUNa}), we see that the diagonal $U(N)$ group acts
by commutation with the ${\Bbb M}$ matrices, whatever their behaviour by
$\cal P$;
the ${\Bbb M}$'s lie in the adjoint representation of this diagonal $U(N)$.

At the level of the group, let ${\cal U}_L \times {\cal U}_R$ be a finite
transformation of the chiral group; we have
\bea
{\cal U}_L \times {\cal U}_R \,.\, {\Bbb M}_{{\cal P}even}  &=&
            {\cal U}_L^{-1}  \,{\Bbb M}_{{\cal P}even} \,{\cal U}_R, \cr
{\cal U}_L \times {\cal U}_R \,.\, {\Bbb M}_{{\cal P}odd}  &=&
            {\cal U}_R^{-1} \,{\Bbb M}_{{\cal P}odd} \,{\cal U}_L,
\label{eq:actionUNg}
\eea
reminiscent of the group action in a $\sigma$-model \cite{CurrentAlgebra}
with a $U(N)_L \times U(N)_R$ group of symmetry.
Note that  ``left'' and ``right'' are swapped in the
action on the $\cal P$-odd scalars with respect to the $\cal P$-even ones..

The expressions in terms of commutators ($[\ ,\ ]$) and anticommutators
($\{\ ,\ \}$) have been kept in eq.~(\ref{eq:actionUNa}) for the reader to
make an easy link with scalars as bound states of fermions (quarks).
Indeed, the reader can easily recover the same expressions for the action of
the chiral group by sandwiching the matrices $\Bbb M$ between
a $N$-vector $\Psi$ of ``quarks'' in the fundamental representation of
$U(N)$,
an its conjugate $\ol\Psi$, and by introducing a $\gamma_5$ in the
definition of all ${\Bbb P}$ pseudoscalar states. The ``left'' and ``right''
generators are then  respectively given a 
$(1-\gamma_5)/2$ or a  $(1+\gamma_5)/2$
projectors when acting on the fermions, and  the laws of transformations of
the latter induce those of the mesons (see \cite{Machet1,Machet2});\l
{\em All group actions on $J=0$ fields written in this work can be
uniquely and straightforwardly deduced from the action on fermions when the
former are written as scalar or pseudoscalar diquark operators.}

\section{The group \boldmath{$SU(2)_L \times U(1)$}.}

The group $SU(2)_L\times U(1)$ of electroweak interactions has, as 
shown below, the fundamental property and advantage that, for $N$ even,
it is a subgroup of the chiral group $U(N)_L \times U(N)_R$. Its generators 
can thus also be taken as $N \times N$ matrices.

The generators of the ``generic'' $SU(2)$, where ``generic'' means
``aligned'' with the chiral group, we take to be
\be
{\Bbb T}^3 = {1\over 2}\left(\begin{array}{ccc}
                        {\Bbb I} & \vline & 0\\
                        \hline
                        0 & \vline & -{\Bbb I}           \end{array}\right),\ 
{\Bbb T}^+ =           \left(\begin{array}{ccc}
                        0 & \vline & {\Bbb I}\\
                        \hline
                        0 & \vline & 0           \end{array}\right),\ 
{\Bbb T}^- =           \left(\begin{array}{ccc}
                        0 & \vline & 0\\
                        \hline
                        {\Bbb I} & \vline & 0           \end{array}\right).
\label{eq:generic}
\ee
The ${\Bbb I}$'s in eq.~(\ref{eq:generic}) stand for $N/2 \times N/2$ unit
matrices (we require ${\Bbb T}^- = ({\Bbb T}^+)^\dagger$, such that the 
unit matrices are chosen to have the same dimension). 
${\Bbb T}^+$ and ${\Bbb T}^-$ are respectively $({\Bbb T}^1 +
i~{\Bbb T}^2)$ and $({\Bbb T}^1 - i~{\Bbb T}^2)$.
A ``left'' generic $SU(2)$ is defined accordingly.

The $U(1)$ of hypercharge, non-diagonal, but which commutes with $SU(2)_L$ 
is defined by its generator $({\Bbb Y}_L, {\Bbb Y}_R)$, with
\bea
{\Bbb Y}_L &=& {1\over 6} \;{\Bbb I},\cr 
{\Bbb Y}_R  &=& \hskip 2.5mm           {\Bbb Q}_R\, ,
\label{eq:U1}
\eea
where ${\Bbb Q} = ({\Bbb Q}_L , {\Bbb Q}_R)$ is the charge operator and 
$\Bbb I$ is the unit $N \times N$ matrix.

The Gell-Mann-Nishijima relation
\be
{\Bbb Y} = {\Bbb Q} - {\Bbb T}\,^3_L,
\label{eq:GMN}
\ee
to be understood as
\be
({\Bbb Y}_L,{\Bbb Y}_R)= ({\Bbb Q}_L, {\Bbb Q}_R) - ({\Bbb T}\,^3_L, 0),
\label{eq:GMNchiral}\ee
is verified, for its left and right projections, by the definitions
(\ref{eq:generic},\ref{eq:U1}) above when the charge operator $\Bbb Q$ is
diagonal with
\be
{\Bbb Q}_L = {\Bbb Q}_R  =            \left(\begin{array}{ccc}
                               {2/3} & \vline & 0\\
                               \hline
                               0 & \vline & {-1/3}           \end{array}\right).
\label{eq:charge}
\ee
The ``alignment'' of the electroweak subgroup inside the chiral group is
controlled by a unitary matrix, $({\Bbb R},{\Bbb R})$, acting diagonally,
with 
\be
{\Bbb R} =             \left(\begin{array}{ccc}
                        {\Bbb I} & \vline & 0\\
                        \hline
                        0 & \vline & {\Bbb K}           \end{array}\right),
\label{eq:rotation}
\ee
where $\Bbb K$ is a $N/2 \times N/2$ unitary matrix of rotation
\cite{Cabibbo,KobayashiMaskawa}.
The ``rotated'' electroweak group is then the one with generators
\be
{\Bbb R}^\dagger {\Bbb T}\; {\Bbb R}\; ;
\label{eq:rotgroup}
\ee
In practice, this rotation only acts on the ${\Bbb T}^\pm$ generators;
explicitly, one has
\be
{\Bbb T}^3_{rotated} = {1\over 2}\left(\begin{array}{ccc}
                        {\Bbb I} & \vline & 0\\
                        \hline
                        0 & \vline & -{\Bbb I}
\end{array}\right),\
{\Bbb T}^+_{rotated} =           \left(\begin{array}{ccc}
                        0 & \vline & {\Bbb K}\\
                        \hline
                        0 & \vline & 0           \end{array}\right),\
{\Bbb T}^-_{rotated} =           \left(\begin{array}{ccc}
                        0 & \vline & 0\\
                        \hline
                        {\Bbb K}^\dagger & \vline & 0    \end{array}\right);
\label{eq:rotated}
\ee

the reader will recognize in eq.~(\ref{eq:rotated}) the usual $SU(2)_L$
generators of the Glashow-Salam-Weinberg model when acting on $N$-vectors of
quarks in the fundamental representation of $U(N)$.

\section{Quadruplet scalar representations of \boldmath{$SU(2)_L \times U(1)$}.}

Because electroweak interactions violate parity, the representations of 
the corresponding group of symmetry mix states of different parities, `scalars' 
and `pseudoscalars'. 
The representations are of two types, $\cal P$-even and $\cal P$-odd, 
according to their transformation properties by the parity changing operator
$\cal P$ already mentioned in section \ref{sec:CHIRAL}.

In the same way (see eq.~(\ref{eq:actionUNa})) as we wrote the action of the 
chiral group on scalar fields represented by $N \times N$ matrices $\Bbb M$,
we define the action of its $SU(2)_L$ subgroup, to which we add the action of
the electric charge $\Bbb Q$ according to:
\be
{\Bbb Q}\,.\,{\Bbb M} =  [{\Bbb M},{\Bbb Q}];
\label{eq:charge1}
\ee
it acts by commutation because it is a diagonal operator (see section 
\ref{sec:CHIRAL}).

We shall now build a very special type of representations of the ``generic''
$SU(2)_L \times U(1)$ group defined in eqs.~(\ref{eq:generic},\ref{eq:U1}).
We write them in the form $({\Bbb M}^0, \vec {\Bbb M})$, where the $\Bbb M$'s
are still $N \times N$ matrices;
$\vec {\Bbb M}$ stands for the sets of complex matrices
 $\{{\Bbb M}^1, {\Bbb M}^2, {\Bbb M}^3\}$ or
$\{{\Bbb M}^3, {\Bbb M}^+, {\Bbb M}^-\}$ with ${\Bbb M}^+ = ({\Bbb M}^1 + i\,
{\Bbb M}^2)/\sqrt{2}\;,\; {\Bbb M}^- = ({\Bbb M}^1 - i\, {\Bbb M}^2)/\sqrt{2}$.

Let us consider quadruplets of the form
\be
({\Bbb M}\,^0, {\Bbb M}^3, {\Bbb M}^+, {\Bbb M}^-) = \left[
 {1\over \sqrt{2}}\left(\begin{array}{ccc}
                        {\Bbb D} & \vline & 0\\
                        \hline
                        0 & \vline & {\Bbb D}           \end{array}\right),
{i\over \sqrt{2}} \left(\begin{array}{ccc}
                        {\Bbb D} & \vline & 0\\
                        \hline
                        0 & \vline & -{\Bbb D}           \end{array}\right),
 i\left(\begin{array}{ccc}
                        0 & \vline & {\Bbb D}\\
                        \hline
                        0 & \vline & 0           \end{array}\right),
 i\left(\begin{array}{ccc}
                        0 & \vline & 0\\
                        \hline
                        {\Bbb D} & \vline & 0        \end{array}\right)
             \right],
\label{eq:reps}
\ee
where $\Bbb D$ is a real $N/2 \times N/2$ matrix. 

The action of $SU(2)_L \times U(1)$ on these quadruplets is defined by its
action on each of the four components, as written  in
eqs.~(\ref{eq:actionUNa},\ref{eq:charge1}).
It turns out that it can be rewritten in the form
(the Latin indices $i,j,k$ run from $1$ to $3$):
\bea
{\Bbb T}^i_L\,.\,{\Bbb M}^j_{{\cal P}even} &=& -{i\over 2}\left(
              \epsilon_{ijk} {\Bbb M}^k_{{\cal P}even} + 
                           \delta_{ij} {\Bbb M}_{{\cal P}even}^0
                              \right),\cr
{\Bbb T}^i_L\,.\,{\Bbb M}_{{\cal P}even}^0 &=& 
                                {i\over 2}\; {\Bbb M}_{{\cal P}even}^i;
\label{eq:actioneven}
\eea
and 
\bea
{\Bbb T}^i_L\,.\,{\Bbb M}_{{\cal P}odd}^j &=& -{i\over 2}\left(
                   \epsilon_{ijk} {\Bbb M}_{{\cal P}odd}^k - 
                           \delta_{ij} {\Bbb M}_{{\cal P}odd}^0
                              \right),\cr
{\Bbb T}^i_L\,.\,{\Bbb M}_{{\cal P}odd}^0 &=& 
                        \hskip 5mm  -{i\over 2}\; {\Bbb M}_{{\cal P}odd}^i.
\label{eq:actionodd}
\eea
The charge operator acts indifferently on ${\cal P}$-even and ${\cal P}$-odd
matrices by:
\bea
{\Bbb Q}\,.\,{\Bbb M}\,^i &=& -i\,\epsilon_{ij3} {\Bbb M}\,^j,\cr
{\Bbb Q}\,.\,{\Bbb M}\,^0 &=& 0\,,
\label{eq:chargeaction}
\eea
and the action of the $U(1)$ generator $\Bbb Y$ follows from eq.~(\ref{eq:GMN}).

Still as a consequence of (\ref{eq:actionUNa}), the action of the 
``right'' group $SU(2)_R$ is of the same form as displayed in 
eqs.~(\ref{eq:actioneven},\ref{eq:actionodd}) but with the signs in front of
${\Bbb M}^0$'s all swapped. 

We see that we deal now with 4-dimensional representations of $SU(2)_L
\times U(1)$, and which are also, by the above remark, representations of
$SU(2)_R$.
In the basis of any such representation, the generators of the electroweak
group can be rewritten as $4 \times 4$ matrices. This is also the case for
the generators of the diagonal  $SU(2)$ (see section \ref{sec:quantization}).

We shall restrict below to this type of representations (\ref{eq:reps}).

They decompose into ``symmetric'' representations, corresponding to 
$\; \Bbb D = \,{\Bbb D}^\dagger$, and ``antisymmetric'' ones for which 
$\; \Bbb D = -\,{\Bbb D}^\dagger$.

There are $N/2(N/2 +1)/2$ independent real symmetric $\Bbb D$ matrices;
hence, the sets of ``even'' and ``odd'' symmetric quadruplet representations 
of the type (\ref{eq:reps}) both have dimension $N/2(N/2 +1)/2$.
Similarly, the antisymmetric ones form two sets of dimension $N/2(N/2 -1)/2$.

If $({\Bbb M}\,^0, \vec {\Bbb M})$ is a representation of the ``generic'' 
$SU(2)_L \times U(1)$ of eqs.~(\ref{eq:generic},\ref{eq:U1}), 
$({\Bbb R}^\dagger {\Bbb M}\,^0 {\Bbb R}, {\Bbb R}^\dagger 
\vec {\Bbb M} {\Bbb R})$ is a representation of the ``rotated'' group of 
eqs.~(\ref{eq:rotgroup},\ref{eq:rotated}); it is called hereafter 
a ``rotated'' representation. It writes explicitly:
\bea
& & ({\Bbb M}\,^0, {\Bbb M}^3, {\Bbb M}^+, {\Bbb M}^-)_{rotated}\cr = 
& & \left[
 {1\over \sqrt{2}}\left(\begin{array}{ccc}
                        {\Bbb D} & \vline & 0\\
                        \hline
                        0 & \vline & {\Bbb K}^\dagger\,{\Bbb D}\,{\Bbb K}
                   \end{array}\right),
{i\over \sqrt{2}} \left(\begin{array}{ccc}
                        {\Bbb D} & \vline & 0\\
                        \hline
                        0 & \vline & -{\Bbb K}^\dagger\,{\Bbb D}\,{\Bbb K}
                   \end{array}\right),
i\left(\begin{array}{ccc}
                        0 & \vline & {\Bbb D}\,{\Bbb K}\\
                        \hline
                        0 & \vline & 0           \end{array}\right),
 i\left(\begin{array}{ccc}
                        0 & \vline & 0\\
                        \hline
                        {\Bbb K}^\dagger\,{\Bbb D} & \vline & 0   
                    \end{array}\right)
             \right].\cr
& &
\label{eq:rotreps}
\eea
Every representation above is a reducible representation of $SU(2)_L$ (or
$SU(2)_R$)
and is the sum of two (complex) representations of spin $1/2$. 
This makes it isomorphic to the standard scalar set of the
Glashow-Salam-Weinberg model \cite{GlashowSalamWeinberg}.

Now, if we consider the transformation properties by the diagonal $SU(2)$,
all $\vec {\Bbb M}$'s are (spin $1$) triplets, lying in the adjoint 
representation, while all ${\Bbb M}^0$'s are singlets. 

To ease the link with physics, let us make one more step in the reshuffling
of our quadruplets.
By summing or subtracting the two representations, ${\cal P}$-even and
${\cal P}$-odd, corresponding to the same set of four $\Bbb M$ matrices, one
can form two other representations; the action of $SU(2)_L$ rewrites, using
eqs.~(\ref{eq:actioneven},\ref{eq:actionodd}):
\bea
{\Bbb T}^i_L\,.\,({\Bbb M}_{{\cal P}even}^j + {\Bbb M}_{{\cal P}odd}^j) &=& 
-{i\over 2}\left(
 \epsilon_{ijk} ({\Bbb M}_{{\cal P}even}^k +{\Bbb M}_{{\cal P}odd}^k) + 
   \delta_{ij} ({\Bbb M}_{{\cal P}even}^0 - {\Bbb M}_{{\cal P}odd}^0)
                              \right),\cr
{\Bbb T}^i_L\,.\,({\Bbb M}_{{\cal P}even}^0 + {\Bbb M}_{{\cal P}odd}^0) &=& 
   {i\over 2}\; ({\Bbb M}_{{\cal P}even}^i - {\Bbb M}_{{\cal P}odd}^i);
\label{eq:actioneven+odd}
\eea
\bea
{\Bbb T}^i_L\,.\,({\Bbb M}_{{\cal P}even}^j - {\Bbb M}_{{\cal P}odd}^j)
&=& 
-{i\over 2}\left(
 \epsilon_{ijk} ({\Bbb M}_{{\cal P}even}^k -{\Bbb M}_{{\cal P}odd}^k) +
   \delta_{ij} ({\Bbb M}_{{\cal P}even}^0 + {\Bbb M}_{{\cal P}odd}^0)
                              \right),\cr
{\Bbb T}^i_L\,.\,({\Bbb M}_{{\cal P}even}^0 - {\Bbb M}_{{\cal P}odd}^0)
&=&
   {i\over 2}\; ({\Bbb M}_{{\cal P}even}^i + {\Bbb M}_{{\cal P}odd}^i).
\label{eq:actioneven-odd}
\eea
As usual, the action of $SU(2)_R$ is obtained from the one above by swapping
the signs of all ${\Bbb M}^0$'s.

It is convenient to rewrite
\be
({\Bbb M}_{{\cal P}even} + {\Bbb M}_{{\cal P}odd}) = {\Bbb S},
\label{eq:scalar}
\ee
and
\be
({\Bbb M}_{{\cal P}even} - {\Bbb M}_{{\cal P}odd}) = {\Bbb P},
\label{eq:pseudo}
\ee
eq.~(\ref{eq:scalar}) corresponding to a scalar state ${\Bbb S}$, and
eq.~(\ref{eq:pseudo}) to a pseudoscalar state ${\Bbb P}$.
Thus, of those two new representations, the first is of the type
\be
({\Bbb M}\,^0, \vec {\Bbb M}) = ({\Bbb S}^0, \vec {\Bbb P})
\label{eq:SP}
\ee
and the second of the type 
\be
({\Bbb M}\,^0, \vec {\Bbb M}) = ({\Bbb P}\,^0, \vec {\Bbb S});
\label{eq:PS}
\ee
both have scalar and pseudoscalar entries, each entry having a definite $P$
{\em quantum number} (we attribute to scalars the parity $P = +1$ and
to pseudoscalars the parity $P = -1$). 
Among the ``symmetric'' $({\Bbb S}^0, \vec {\Bbb P})$ representations lies
the one
corresponding to ${\Bbb D} = {\Bbb I}$ and which thus includes the scalar
$U(N)$ singlet: it is hereafter identified with the Higgs boson $H$ and the
corresponding representation with the usual scalar 4-plet of the
Standard Model:
\be
(H, \vec \phi) =
 \left[   {1\over \sqrt{2}} \left(\begin{array}{ccc}
                        {\Bbb I} & \vline & 0\\
                        \hline
                        0 & \vline & {\Bbb I}
                   \end{array}\right)_{\Bbb S},
          {i\over \sqrt{2}} \left(\begin{array}{ccc}
                        {\Bbb I} & \vline & 0\\
                        \hline
                        0 & \vline & -{\Bbb I}
                   \end{array}\right)_{\Bbb P},
          i\left(\begin{array}{ccc}
                        0 & \vline & {\Bbb K}\\
                        \hline
                        0 & \vline & 0           
                   \end{array}\right)_{\Bbb P},
           i\left(\begin{array}{ccc}
                        0 & \vline & 0\\
                        \hline
                        {\Bbb K}^\dagger & \vline & 0
                    \end{array}\right)_{\Bbb P}
             \right] .
\label{eq:goldstone}
\ee
{From} now onwards, we shall work with the representations (\ref{eq:SP}) and
(\ref{eq:PS}).
By hermitian conjugation a ``symmetric'' $({\Bbb M}\,^0, \vec {\Bbb M})$ 
representation gives $({\Bbb M}\,^0, -\vec {\Bbb M})$; an ``antisymmetric'' 
representation gives $(-{\Bbb M}\,^0, \vec {\Bbb M})$; the representations
(\ref{eq:SP}) and (\ref{eq:PS}) are consequently
representations of given $CP$\ (charge\ conjugation\ $\times$\  parity):
``symmetric'' $({\Bbb S}^0, \vec {\Bbb P})$'s and ``antisymmetric'' 
$({\Bbb P}\,^0, \vec {\Bbb S})$'s are $CP$-even, while ``symmetric'' 
$({\Bbb P}\,^0, \vec {\Bbb S})$'s and ``antisymmetric'' 
$({\Bbb S}^0, \vec {\Bbb P})$'s are $CP$-odd.
By multiplying a representation of a given $CP$ by $i$, one obtains a
representation with opposite $C$, and thus with opposite $CP$; they however
correspond to the same Lagrangian (see eq.~(\ref{eq:lagrangian}) below).

\section{The \boldmath{$SU(2)_L \times U(1)$} invariant Lagrangian for 
scalar fields.}

To every representation (\ref{eq:reps}), in particular those of the form
(\ref{eq:SP}) or (\ref{eq:PS}), is associated a unique quadratic 
expression invariant by any $SU(2)_L\times U(1)$ transformation:
\be
{\cal I} = {\Bbb M}\,^0 \otimes {\Bbb M}\,^0 + 
                 \vec {\Bbb M} \otimes \vec {\Bbb M};
\label{eq:invar}
\ee
the ``$\otimes$'' product is a tensor product; it is {\em not} meant in the 
sense of the usual 
multiplication of matrices but in the sense of the product of fields as 
functions of
space-time. $\vec {\Bbb M} \otimes \vec {\Bbb M}$ stands for 
$\sum_{i=1,2,3} {\Bbb M}\,^i \otimes  {\Bbb M}\,^i$.

The invariant $\cal I$ of eq.~(\ref{eq:invar}) is also invariant by
$SU(2)_R$.

Once we have the action of the (gauge) group and a quadratic invariant, we can
immediately write a gauge invariant electroweak Lagrangian for the $2N^2$
scalar and pseudoscalar fields. If we do not allow for scalar-pseudoscalar
transitions, it includes {\em a priori} $N^2/2$ independent electroweak mass 
scales, one for each quadruplet (we thus exclude mass terms 
proportional to the second type of possible quadratic invariant, linking one
quadruplet of the form (\ref{eq:SP}) with one of the form (\ref{eq:PS}):
${\Bbb S}^0 \otimes {\Bbb P}^0 + \vec{\Bbb P} \otimes \vec{\Bbb S}$.)
\be
{\cal L} = \sum_{all\ reps\ {\cal R}}{1\over 2} \left(
D_\mu {\Bbb M}_{\cal R}^0 \otimes D^\mu {\Bbb M}_{\cal R}^0 
    + D_\mu \vec {\Bbb M}_{\cal R} \otimes D^\mu \vec {\Bbb M}_{\cal R}
- m_{\cal R}^2 ({\Bbb M}_{\cal R}^0 \otimes {\Bbb M}_{\cal R}^0 +
                 \vec {\Bbb M}_{\cal R} \otimes \vec {\Bbb M}_{\cal R})
                                 \right),
\label{eq:lagrangian}
\ee
where the sum is extended to all representations (\ref{eq:SP}) and
(\ref{eq:PS}).

$D_\mu$ in eq.~(\ref{eq:lagrangian}) is the covariant derivative with
respect to $SU(2)_L \times U(1)$:
\be
D_\mu {\Bbb M}^\alpha = \p_\mu {\Bbb M}^\alpha 
                  -i g' B_\mu\;{\Bbb Y}.{\Bbb M}^\alpha
           -ig  (W_\mu)_i {\Bbb T}^i_L.{\Bbb M}^\alpha,
\label{eq:covder}
\ee
where $g'$ and $g$ are respectively the weak hypercharge and $SU(2)_L$ coupling
constants, and $B_\mu$ and $\vec W_\mu$ the associated gauge fields.
The explicit expressions for the covariant derivatives of the scalar fields
can be found in eqs.~(\ref{eq:covders}) below.

The link with a Lorentz scalar for the Lagrangian density above goes along
the following simple lines:
every matrix $\Bbb M$ in a quadruplet, describing an electroweak state, can
be expanded on the basis of strong eigenstates, themselves $N\times N$
matrices; those are then replaced in $\cal L$ by the corresponding mesonic
fields like $\pi^+, K^0 \ldots $, and the tensor product by the usual
multiplication of scalars. The obtained new Lagrangian density is now a
Lorentz scalar. It is of course important that the kinetic terms can be
diagonalized both in the electroweak and in the strong basis. Then, if they
are all normalized to $1$, the Lagrangian (\ref{eq:lagrangian}) has, 
at the limit $g,g' \rightarrow 0$, a global $SU(2)_R \times SU(2)_L$ symmetry 
which exists independently of the values of the masses $m_{\cal R}$.

The three Goldstones $\vec \phi$, absorbed by the three gauge fields who
become massive, are themselves linear combinations of ``strong''
eigenstates, as for example, in the case $N=4$, we get from
(\ref{eq:goldstone})
\be
\phi^+ = a\left[c_\theta (\pi^+ + D_s^+) + s_\theta (K^+ - D^+)\right],
\ee
where $c_\theta$ and $s_\theta$ are the cosine and sinus of the Cabibbo
angle, and $a$ a scaling factor 
\be
a= {f\over \la H \ra},
\ee
($\la H \ra = {v\over \sqrt{2}}$), which has already been studied in 
\cite{Machet1} \cite{Machet2}; $f$ is the
leptonic decay constant supposed to be the same for all concerned mesons.
The link with observed mesons was shown in \cite{Machet1} \cite{Machet2} to
be as follows:\l
- all matter fields, mesons, leptons and gauge fields have to be rescaled by
the factor $a$; for example in the one-generation case
\be
(H, \vec\phi) = a (H', \vec\pi);
\ee
for the leptons
\be
\psi_\ell = a \psi'_\ell,
\ee
and for the gauge field $\sigma_\mu$, ($\sigma_\mu =\vec W_\mu, B_\mu$)
\be
\sigma_\mu = a \sigma'_\mu;
\ee
- all coupling constants have to be rescaled by $1/a$, generically for
$\kappa = g, g'$
\be
\kappa = \kappa'/a.
\ee
The Lagrangian for all rescaled matter fields  to be considered is then 
${\cal L}'$ = ${\cal L}/a^2$, which yields the usual leptonic and semi-leptonic
amplitudes for the pions $\vec\pi$ to decay into ``primed'' leptons 
$\psi'_\ell$, as given  by the ``Partially Conserved Axial Current'' 
hypothesis; the rescaled fields $\vec \pi, \psi'_\ell, \sigma'_\mu$  are 
considered to be the physical fields, which interact with the physical 
coupling constants $\kappa'$.

One avoids in this way the mass scale problem occurring in theories with 
dynamical symmetry breaking and the necessity of introducing a new scale of 
interaction with associated super-heavy mesons, like in technicolour theories
\cite{SusskindWeinberg}. 

A point to be stressed here is that this Lagrangian includes only
renormalizable couplings. We have in addition the freedom to add all
biquadratic terms, which will automatically respect the local $SU(2)_L
\times U(1)$ symmetry.
It is furthermore anomaly-free, no anomaly arising from scalar fields
\cite{AdlerBellJackiw}.

It has additional symmetry properties, to which we now turn.

\section{The \boldmath{$SU(2)_V$} ``custodial'' symmetry.}

The 4-dimensional representations (\ref{eq:reps}) of $SU(2)_L \times U(1)$
have already been mentioned to be representations of $SU(2)_R$. They are
thus naturally representations of the diagonal $SU(2)_V$, that we study in
more detail.

When acting in the 4-dimensional vector space of which (\ref{eq:reps}) form 
a basis, its generators ${\Bbb T}^3, {\Bbb T}^\pm$ can be represented
as $4\times 4$ matrices $\tilde T^3, \tilde T^\pm$; explicitly:
\be
\tilde{\Bbb T}^+ = \left( \ba{cccc}
                   0  &  0        &  0       &  0  \cr
                   0  &  0        & \sqrt{2} &  0  \cr
                   0  &  0        &  0       &  0  \cr
                   0  & -\sqrt{2} &  0       &  0
\ea\right),\quad
\tilde{\Bbb T}^- = \left( \ba{cccc}
                   0  &  0        &  0       &  0   \cr
                   0  &  0        &  0       & -\sqrt{2} \cr
                   0  &  \sqrt{2} &  0       &  0 \cr
                   0  &  0        &  0       &  0
\ea\right),\quad
\tilde{\Bbb T}^3 = \left( \ba{cccc}
                   0  &  0        &  0       &  0  \cr
                   0  &  0        &  0       &  0  \cr
                   0  &  0        & -1       &  0  \cr
                   0  &  0        &  0       &  1
\ea\right).
\label{eq:SU2V}\ee
That the first line in any of the three above matrices identically vanishes
is the translation of the already mentioned fact that the first entry 
${\Bbb M}^0$ of the representations (\ref{eq:reps}) are singlets by the 
diagonal $SU(2)$, while the three other entries $\vec{\Bbb M}$ form a 
triplet in the adjoint representation.

We show now that the whole Lagrangian (\ref{eq:lagrangian}) has a global
$SU(2)_V$ symmetry, when the gauge fields 
$W_\mu^\pm$ and $ \tilde Z_\mu = Z_\mu/\cos\theta_w$, 
with $\theta_w$ the Weinberg angle,
transform like a vector in the adjoint representation of $SU(2)_V$. This is
not a surprise since those precisely absorb the $\vec \phi$ triplet of
eq.~(\ref{eq:goldstone}), also in
the adjoint, to become massive, when the gauge symmetry is broken down from
$SU(2)_L \times U(1)_{\Bbb Y}$ to $U(1)_{em}$. The normalization of the last 
one ensures that the resulting mass term for the gauge fields
$M_W^2 (2 W_\mu^+ W^{\mu -} + Z_\mu Z^\mu/c_W^2)$ satisfies $\rho =1$, where
$\rho = M_W/(M_Z \cos\theta_w)$ is the Michel's parameter. 
We recover the well-known link between the 
custodial $SU(2)_V$ and the value of $\rho$ \cite{Sikivie}.\l
For this purpose, let us explicitly write the covariant
(with respect to $SU(2)_L \times U(1)$) derivatives of a quadruplet, and
show that they transform like a singlet plus a triplet by the custodial
$SU(2)$. We do it explicitly for a $\cal P$-even quadruplet.

\vbox{
\bea
D_\mu {\Bbb M}^0_{even} &=& \p_\mu {\Bbb M}^0_{even} + {e\over 2 s_w} 
                   (W_\mu^1{\Bbb M}^1_{even}  + W_\mu^2{\Bbb M}^2_{even} 
                    + (Z_\mu/c_w){\Bbb M}^3_{even} ),\cr
     &=& {\cal D}_\mu {\Bbb M}^0_{even} + {e\over 2 s_w} 
                  (W_\mu^1{\Bbb M}^1_{even}  + W_\mu^2{\Bbb M}^2_{even}
                    + (Z_\mu/c_w){\Bbb M}^3_{even} ),\cr
D_\mu {\Bbb M}^3_{even} &=& \p_\mu {\Bbb M}^3_{even} + {e\over 2 s_w} \left(
                    i\,(W_\mu ^+ {\Bbb M}^-_{even}  - W_\mu^-{\Bbb M}^+_{even})
                      - (Z_\mu/c_w){\Bbb M}^0_{even} \right),\cr
     &=& {\cal D}_\mu {\Bbb M}^3_{even} 
                        -{e\over 2 s_w}(Z_\mu /c_w){\Bbb M}^0_{even},\cr
D_\mu {\Bbb M}^+_{even} &=& \p_\mu {\Bbb M}^+_{even} -{e\over 2 s_w} \left(
                    W_\mu^+({\Bbb M}^0_{even} + i {\Bbb M}^3_{even})
             -i (Z_\mu/c_w){\Bbb M}^+_{even}\right) 
             +i {e\over c_w}B_\mu {\Bbb M}^+_{even}, \cr
     &=& {\cal D}_\mu {\Bbb M}^+_{even}
                 -{e\over 2 s_w} W_\mu^+{\Bbb M}^0_{even}
                 +i{e\over c_w} B_\mu {\Bbb M}^+_{even},\cr
D_\mu {\Bbb M}^-_{even} &=& \p_\mu {\Bbb M}^-_{even} -{e\over 2 s_w} \left(
                    W_\mu^-({\Bbb M}^0_{even} - i {\Bbb M}^3_{even})
             +i (Z_\mu/c_w){\Bbb M}^-_{even}\right) 
             -i {e\over c_w} B_\mu {\Bbb M}^-,\cr
     &=& {\cal D}_\mu {\Bbb M}^-_{even}
                  -{e\over 2 s_w} W_\mu^-{\Bbb M}^0_{even}
                   -i{e\over c_w} B_\mu {\Bbb M}^-_{even}.
\label{eq:covders}
\eea
}
In eq.~({\ref{eq:covders}) above, we noted $c_w$ and $s_w$ respectively the
cosine and sine of the Weinberg angle.  $A_\mu$ is the photon, $W_\mu^\pm 
= (W_\mu^1 \pm i W_\mu^2) /\sqrt{2}$, and we have as usual
\bea
g &=& {e\over s_w},\  g' = {e\over c_w},\cr
Z_\mu &=& c_w W_\mu^3 - s_w B_\mu,\  A_\mu = c_w B_\mu + s_w W_\mu^3.
\eea
${\cal D}_\mu$ is the covariant derivative with respect to the diagonal
$SU(2)_V$ group
\be
{\cal D}_\mu {\Bbb M} = \p_\mu {\Bbb M}
            -i{e\over s_w}({1\over\sqrt{2}}(W_\mu^+ \tilde{\Bbb T}^-
       + W_\mu^- \tilde{\Bbb T}^+) 
+{Z_\mu \over c_w}\tilde {\Bbb T}^3).{\Bbb M}\ .
\ee
The normal derivative of $\Bbb M$ transforming like $\Bbb M$ itself,
that $D_\mu {\Bbb M}^0$ is a singlet of $SU(2)_V$ is trivial as soon as, as
stressed before, $\vec{\Bbb M}$ is a triplet in the adjoint and $(W_\mu ^\pm,
Z_\mu/c_w)$ too, since the scalar product of those two vectors is an invariant;
\l
that the three other covariant derivatives transform like a vector results from
the three following facts:\l
- from the 2 vectors $\vec{\Bbb M}$ and $(W_\mu ^\pm,Z_\mu/c_w)$ we can form
a third one with the $\epsilon_{ijk}$ tensor
\be \left(\ba{l}
{\Bbb M}^- W_\mu^+    - {\Bbb M}^+ W_\mu^-,\cr
{\Bbb M}^3 W_\mu^+    - {\Bbb M}^+ (Z_\mu/c_w),\cr
{\Bbb M}^3 W_\mu^-    - {\Bbb M}^- (Z_\mu/c_w);
\ea\right)\ee
- ${\Bbb M}^0$ being a singlet by $SU(2)_V$, the terms ${\Bbb M}^0 W_\mu^\pm$ 
transform like $W_\mu^\pm$ and thus like ${\Bbb M}^\pm$, $(Z_\mu /c_w){\Bbb
M}^0$ like $(Z_\mu /c_w)$ and thus like ${\Bbb M}^3$;\l
- $B_\mu$ is to be considered as a singlet of $SU(2)_V$, such that the terms
$(B_\mu /c_w){\Bbb M}^\pm$ transform like ${\Bbb M}^\pm$.

The same argumentation works for $\cal P$-odd scalars. Their covariant
derivatives are immediately obtained from eqs.~(\ref{eq:covders}) above by
changing the signs of all ${\Bbb M}^0$'s.

This shows the existence of a  global $SU(2)_V$ custodial symmetry for
the Lagrangian (\ref{eq:lagrangian}),
independently of the value of the hypercharge coupling $g'$.

The starting global $U(N)_L \times U(N)_R$ symmetry of strong interactions is
broken down to $SU(2)_R \times SU(2)_L$ by the $N^2/2$ electroweak mass
scales introduced in the Lagrangian (\ref{eq:lagrangian}); this symmetry is
only (classically) exact when the coupling constants $g,g' \rightarrow 0$.
When they are turned on, it is broken
down to the custodial $SU(2)_V$; electroweak interactions
are themselves broken down to $U(1)_{em}$, but the
custodial symmetry is preserved, at least classically.

As all relevant $SU(2)$'s are the ``rotated'' ones, in particular, 
the custodial $SU(2)_V$ is not the strong isospin group \cite{Sikivie}.

After symmetry breaking, there exists {\em a priori} two $SU(2)_V$ mass scales
for each quadruplet: the first is that of the vector triplet, the second
that of the singlet, like for example the Higgs boson. So, in this framework,
and without adding more information (like compositeness as has been done in
\cite{Machet1,Machet2}), there exists no link between the mass of the Higgs 
boson and that of the other gauge fields or $J=0$ mesons.

Let us now examine whether this symmetry can be considered as a 
local symmetry.

Making a space-time dependent $SU(2)_V$ transformation with parameters 
$\vec\theta$ on the scalar fields and transforming the vector fields
$W_\mu^\pm, Z_\mu/c_w$ like the corresponding gauge potentials 
($B_\mu$ being a singlet does not transform), 
ones finds from (\ref{eq:covders}) that the Lagrangian
(\ref{eq:lagrangian}) varies, for each quadruplet, by
\be
\Delta{\cal L} = {\cal D}_\mu\vec\theta .(\vec{\Bbb M}\otimes D^\mu {\Bbb M}^0 
                  -{\Bbb M}^0\otimes  D^\mu\vec{\Bbb M}),
\label{eq:deltaL}\ee
such that the existence of a  local custodial $SU(2)_V$ symmetry is
linked to the conservation of the triplet of currents $\vec V^\mu$
\be
{\cal D}_\mu \vec V^\mu = 0,
\ee
with
\be
\vec V^\mu = \vec {\Bbb M}\otimes  D^\mu {\Bbb M}^0 
              - {\Bbb M}^0\otimes  D^\mu \vec{\Bbb M}.
\label{eq:vmu}\ee
$\vec V_\mu$ is an $SU(2)_V$ triplet. Its ``singlet'' partner $V_\mu^0$
identically vanishes by the definition (\ref{eq:vmu}).

These currents are automatically covariantly (with respect to $SU(2)_L
\times U(1)$) conserved by the classical
equations of motion for the $\Bbb M$ fields, as can be seen from
(\ref{eq:vmu}), which entails
\be
D^\mu V^i_\mu = {\Bbb M}^i\otimes  D^2 {\Bbb M}^0 - 
                                     {\Bbb M}^0\otimes  D^2 {\Bbb M}^i,
\ee
and  from the Lagrangian (\ref{eq:lagrangian}) to which we can add any term 
quadratic in the invariants $\cal I$ for any quadruplet.

Now, 
\be
D^\mu V^i_\mu = {\cal D}^\mu V^i_\mu -ig' B_\mu \tilde{\Bbb Q}.V^i_\mu,
\ee
where we have used the Gell-Mann-Nishijima relation and the fact that, 
since $V_\mu ^0$ identically vanishes, the ``left'' $SU(2)_L$ acts on $\vec
V_\mu$ like the diagonal $SU(2)_V$.
 
We can thus conclude that the custodial symmetry, which is a  global
symmetry, becomes  local when the hypercharge coupling $g'$ goes to zero.

A vector-like local symmetry, having no anomaly, is preserved at the
quantum level. At the limit $g' \rightarrow 0$, the custodial $SU(2)_V$ 
symmetry is thus an exact  local
symmetry of the standard $SU(2)_L \times U(1)$ Lagrangian
(\ref{eq:lagrangian}) for $J=0$ fields, with gauge fields $W_\mu^\pm, 
Z_\mu/c_w$.

This is to be compared with non-linear $\sigma$-models built on a gauge group 
${\cal G}/{\cal H}$, where $\cal H$ is the little group of the broken $\cal G$
symmetry 
(for example ${\cal G} = SU(2)_L \times SU(2)_R$ and ${\cal H} = SU(2)_V$),
which possess a hidden $\cal H$ local symmetry \cite{Balachandran}; 
however, there, the gauge 
fields are not dynamical; as these are non-renormalizable theories, some 
authors \cite{Gatto} argue that kinetic terms for the gauge fields are 
generated at higher
orders in a loop expansion, to treat these fields as dynamical, with the
possible outcome of new physics. It appears that, in our approach, which is
renormalizable, no new gauge field springs out in the problem.

The presence of this exact custodial symmetry entails, in particular, 
that all corrections to $\rho = 1$ for the Michel's parameter should vanish 
with $g'$ when one computes the quantum corrections with the electroweak
eigenstates displayed here, and not with quarks; the ``screening'' theorem
\cite{Veltman} becomes exact in this limit, which is also that
where $Z_\mu = W_\mu^3, A_\mu = B_\mu$ and the three massive
vector bosons become degenerate in mass.

\section{Quantization of the electric charge.}\label{sec:quantization}

In the same way as we wrote the $SU(2)_V$ generators as $4\times 4$ matrices
when acting in the 4-dimensional vector spaces spanned by our quadruplets,
we can express the electric charge operator, which turns out to be
\be
\tilde{\Bbb Q} = \left( \ba{cccc}
                   0  &  0        &  0       &  0  \cr
                   0  &  0        &  0       &  0  \cr
                   0  &  0        & -1       &  0  \cr
                   0  &  0        &  0       &  1 
\ea\right).
\label{eq:elec}\ee
By comparison with (\ref{eq:SU2V}), we see that $\tilde Q$ is identical 
with $\tilde T^3$ and that we have the commutation relation
\be
[{\tilde T}^+, {\tilde T}^-] = 2\;{\tilde T}^3 = 2\;{\tilde Q}.
\ee
$\tilde Q$ being an $SU(2)$ generator, its eigenvalues, hence the electric
charges of the representations (\ref{eq:reps}), are quantized like those of
the $z$ component of an angular momentum if the custodial symmetry, that we
exhibited at the classical level, stays an unbroken symmetry of the theory
at the quantum level, as suggested by experimental results (see for example
(\cite{Hollik})).

The $J=1$ mesons naturally fit into triplets and singlets of
both the electroweak $SU(2)_L$ gauge group, and of the
custodial $SU(2)_V$. The two groups act in the same way. 
This is easily seen in a ``composite'' picture where a
vector meson is written as $\rho^\mu = \overline \Psi \gamma^\mu {\Bbb A}
\Psi$, where $\Bbb A$ is again a $U(N)$ generator. The mechanism displayed
above that leads to the quantization of the electric charge is consequently 
even simpler in this case than for the $J=0$ case.

We expect the same mechanism also to operate in the sector of the gauge
fields,  due to their connection to the ``Goldstones'' $\vec \phi$
of the broken symmetry, that they absorb to become massive, 

This general result is to be related with a recent work by Cho and Maison
\cite{ChoMaison} claiming the existence of non trivial classical solutions
of the dyon type in the Glashow-Weinberg-Salam model with one complex
doublet of scalar fields. The point is that, after
extracting the invariant $\cal I$ of the $(H, \vec\phi)$ quadruplet 
(\ref{eq:goldstone}), the gauge group acts on the
remaining real degrees of freedom, living now in a compactified space
$CP^1$; this yields a non-trivial topological structure, at the origin of
the existence of monopole-like solutions of the classical equations of
motion, structure which was thought
before to be only found in a pure $SU(2)$ broken gauge theory with a triplet
of gauge fields and another of scalars, both in the adjoint representation
of the gauge group \cite{Rajaraman}. The Standard Model could consequently
appear as a candidate for a theory where the two possible known ways of having 
a quantization of the electric charge, the presence of ``monopoles'' and the
fact that the generator of the electric charge is one among the generators
of a non-abelian simple group, are two aspects of the same phenomenon,
called electric-magnetic duality \cite{Olive}, without having to invoke
supersymmetry for technical reasons like non-renormalization theorems. 
If this is the case, then one expects the existence, in the
same quantum theory, of a strongly-coupled `magnetic' sector, the particles
of which being solitons or `skyrmion'-like \cite{Skyrmions}.

\section{Particles: a few brief remarks.}\label{sec:PARTICULE}

Some phenomenological aspects have already been tackled in \cite{Machet1,
Machet2}, where the $J=0$ mesons have been explicitly considered as composite.
We recall that the latter approach is totally compatible with the present
one, where fermions do not explicitly appear. So, the reader can refer to the
above works to get more information about leptonic and semi-leptonic decays,
the disappearance of anomalies etc\ldots
We shall only make here some general remarks guiding the connection
with phenomenology.

In the present work, the ``standard'' scalar 4-plet (or complex doublet) of the 
Glashow-Salam-Weinberg model is
identified with the ``symmetric'' $({\Bbb S}^0, \vec {\Bbb P})$ representation 
(\ref{eq:goldstone})
including the scalar $U(N)$ singlet, represented by the $N \times N$ unit 
matrix. The latter is {\em a priori}
chosen as the only diagonal $N \times N$ matrix with a non vanishing trace, 
and so unambiguously defined as the Higgs boson. It is the only field
supposed to have a non-vanishing vacuum expectation value.

Any linear combination of the representations (\ref{eq:SP}) and
(\ref{eq:PS}) also being a
representation, only physical observation can guide us towards the 
determination of what are the observed electroweak eigenstates.
Mixing matrices link physical states with the (rotated) representations
displayed above; they can {\em a priori} depend on new parameters,
differing or not from the angles and phases characterizing, in  the rotation 
matrix $\Bbb R\,$, the alignment of the electroweak group inside the 
chiral group.
 Combining representations of both types $({\Bbb S}^0,\vec {\Bbb P})$ and 
$({\Bbb P}\,^0, \vec {\Bbb S})$ seems however not desired since it would mix
states of different parities. 

Any state produced by strong interactions is a combination of electroweak
eigenstates, which evolve and decay according to the dynamics of electroweak
interactions if no strong channel is allowed for the decay of the 
initial state.

Only those representations of the $({\Bbb S}^0,\vec {\Bbb P})$ type, 
and for which the
scalar entry has a non-vanishing component on the Higgs boson, will undergo
leptonic decays of their three (pseudoscalar) $\vec {\Bbb P}$ entries; 
indeed, only for those
representations will the kinetic term in the Lagrangian (\ref{eq:lagrangian})
include a 
$\la H \ra \sigma^\mu \p_\mu {\Bbb P}$ coupling, where $\sigma^\mu$ is a gauge
field; the direct coupling of the latter to leptons will trigger the
leptonic decay of the pseudoscalar ${\Bbb P}$. In the same way, we deduce
that, by our hypotheses, scalar mesons never leptonically decay: 
by the action of the group, a scalar is connected either to a pseudoscalar,
which is supposed to have a vanishing vacuum expectation value,  
or to another scalar; but the latter is always one with a vanishing
vacuum expectation value since the Higgs boson can only be reached
by acting with the group on a pseudoscalar.

Semi-leptonic decays between states of the same parity can only occur
between the members of  the (diagonal) $SU(2)$ triplet of a given
quadruplet, since the gauge group only connects the entries of a
given representation; indeed, the kinetic term includes couplings of the type
$ {\Bbb P}_1 \sigma^\mu \p_\mu {\Bbb P}_2$, with the gauge field $\sigma_\mu$ 
giving leptons as before. In particular, a (diagonal) $SU(2)$ singlet like a 
${\Bbb P}\,^0$ or a ${\Bbb S}^0$ never semi-leptonically decays into another 
meson of the same parity.

The customary attribution of $CP$ quantum numbers and the presence or not 
of semi-leptonic decays makes that the ``short-lived'' neutral kaon, which
is not observed to decay semi-leptonically, is likely
the $SU(2)$``singlet'' of an ``antisymmetric'' $({\Bbb P}\,^0, \vec {\Bbb S})$ 
representation $(CP = +1)$, or of $i \times$ a ``symmetric''
$({\Bbb P}\,^0, \vec {\Bbb S})$ representation, which has the same $CP$,
 while the neutral 
pion, if thought of as aligned with the corresponding ``strong''
eigenstate, and the ``long-lived'' kaon should {\em a priori} be looked for 
in ``antisymmetric'' $({\Bbb S}^0, \vec {\Bbb P})$ representations $(CP = -1)$. 
or in $i \times$ ``symmetric'' $({\Bbb S}^0, \vec {\Bbb P})$
representations. 
The study of $CP$ violation, in particular the decays of $K$ mesons into two
or three pions is worth a special study, which we postpone to a separate
work.

\section{Conclusion. Outlook.}

Because difficult steps, necessary to go from fundamental fields in the
Lagrangian, like quarks, to observed physics 
are not yet mastered and the formidable problem of confinement unsolved,
we thought that the goal of building a renormalizable gauge 
theory for the interactions of observed particles like mesons was worth
considering.

The custodial symmetry that we exhibited in this model is linked to the 
quantization of the
electric charge, phenomenon usually looked for in unified theories based on a
non-abelian simple gauge group like $SU(5)$ \cite{GeorgiGlashow}, 
or in theories which possess
non-trivial classical solutions with a magnetic charge.

That both mechanisms for charge quantization might occur in the
Glashow-Salam-Weinberg model  with one (or several like here) complex
doublet of scalars \cite{ChoMaison} suggests a much richer content of 
the underlying
quantum theory, with two different phases, the small-coupling (``electric'') 
one corresponding to the usual action that we studied here, and another 
(``magnetic'') one corresponding to  a
strongly-coupled sector, where fundamental particles would be of the
soliton-type. It is natural to think, in agreement with the ideas pioneered by
Skyrme \cite{Skyrmions} and later developed by Witten \cite{Skyrmions}, 
that this second sector could have something to do with baryons,
and/or in general with strong interactions of hadrons.
This will be one of the directions of future works.

\bigskip
\begin{em}
\underline {Acknowledgments}: it is a pleasure to thank my colleagues at
LPTHE, specially O. Babelon, and G. Thompson for fruitful discussions and 
comments.
\end{em}
\newpage\null
\begin{em}

\end{em}

\end{document}